 \documentclass[preprint2]{aastex}

\slugcomment{}

\shorttitle{Dust opacity in normal galaxies}
\shortauthors{Farrah et al.}

\begin{document}


\title{Sub-millimeter emission from type Ia supernova host galaxies at $z=0.5$}


\author{D. Farrah}
\affil{SIRTF Science Centre, California Institute of Technology, Jet Propulsion Laboratory, Pasadena, CA 91125, USA}

\author{M. Fox, M. Rowan-Robinson, and D. Clements}
\affil{Astrophysics Group, Blackett Laboratory, Imperial College, Prince Consort Road, London SW7 2BW, UK}
\and
\author{J. Afonso}
\affil{CAAUL, Observat\'{o}rio Astron\'{o}mico de Lisboa, Tapada da Ajuda, 1349-018 Lisboa, Portugal}


\begin{abstract}
We present deep sub-millimetre observations of seventeen galaxies at $z=0.5$, selected through 
being hosts of a type 1a supernova. Two galaxies are detected directly, and the sample is 
detected statistically with a mean $850\mu$m flux of 1.01mJy$\pm0.33$mJy, which is $25\% - 135\%$ 
higher than locally. We infer that the mean value of $A_{V}$ in normal galaxies at $z=0.5$ 
is comparable to or greater than the mean $A_{V}$ in local normal galaxies, in agreement 
with galaxy chemical evolution models and indirect observational evidence. Scaling from the 
local value given by \citet{rr2} gives a mean extinction at $z=0.5$ of 
$\langle A_{V} \rangle = 0.56\pm0.17$. The dust in the brightest sub-mm object in our sample 
is best interpreted as normal `cirrus' dust similar to that seen locally. The detection rate 
of our sample suggests that some sources found in blank-field sub-mm surveys may not be high 
redshift starbursts, but rather cirrus galaxies at moderate redshifts and with lower star 
formation rates. Finally, an increase in host dust extinction with redshift may impact the 
cosmological results from distant supernova searches. This emphasizes the need to carefully 
monitor dust extinction when using type Ia supernovae to measure the cosmological parameters. 
\end{abstract}


\section{Introduction}
In recent years, deep surveys have found galaxies and QSOs at redshifts greater 
than six, making it possible to trace the cosmic history of star formation and 
AGN activity from the epoch of reionization to the present day. In practice however, 
achieving these goals is difficult, as translating observations of light from 
galaxies into measures of stellar content and formation histories is affected by 
the obscuring effects of dust. Whilst dust obscuration is known to have a significant 
effect locally (e.g. \citet{tma}), the role of dust at higher redshifts is even more 
important. Chemical evolution models \citep{cal,pei} predict that dust obscuration 
should peak at 2-3 times the current value at $1<z<2$, before slowly declining. 
Observationally, the importance of dust obscuration at high redshifts is exemplified 
by the very different cosmic star formation histories inferred from optical and 
infrared surveys. Infrared surveys produce star formation histories that rise more 
sharply from $z=0$ to $z=1$, and show a slower decline at higher redshifts, than 
the star formation histories inferred from optical surveys \citep{lil,mad,ste,rr0,hug}.

These surveys however do not examine extinction evolution in inactive 
galaxies. Blank-field infrared surveys are biased towards systems containing 
a dusty starburst or AGN, whereas Lyman Break galaxies are selected on rest-frame UV 
flux, which may bias them towards having high rates of unobscured star formation. 
In this paper, we present preliminary results from a program to use deep sub-millimetre 
observations to investigate the evolution of sub-mm emission and dust opacity with redshift in what should 
be normal galaxies; the host galaxies of distant type Ia supernovae. Observations and 
data analysis are described in \S\ref{obs}, and results are presented in \S\ref{res}. 
We discuss mean extinction levels and dust properties in \S\ref{dis}, and also mention 
some implications for deep sub-millimetre surveys and supernova cosmology. We assume 
$\Omega_{0}=1$, $\Omega_{m}=0.3$, $\Lambda=0.7$, and $H_{0}=70$.

\section{Observations} \label{obs}
The hosts of the type Ia supernovae discovered by 
the Supernova Cosmology Project (SCP, \citet{per1}) and the High Redshift Supernova Search 
Team (HZT, \citet{rie1}) are ideal targets for studying the evolution of dust opacity in 
normal galaxies; the galaxies have spectroscopic 
redshifts, the galaxies are selected independent of their apparent magnitudes as the 
discovery is based on the supernova rather than the host, and type Ia supernovae come 
from evolved stellar populations. Furthermore, since the supernova spectrum 
must be disentangled from the host galaxy spectrum, none of the host galaxies contain 
the signature bright emission lines of a starburst or AGN (\citet{rie1,per1}, A Riess 
priv. comm., G Aldering priv. comm.).  

We observed seventeen host galaxies of supernovae discovered 
by the SCP and the HZT. The only selection criteria were that the host 
galaxy redshifts should lie in a narrow interval around  z=0.5, so that the fluxes could 
be coadded without $k$-correcting to a common wavelength, and that the supernova in the host 
should be spectroscopically confirmed as a type Ia.  
The host galaxies were observed at $450\mu$m and $850\mu$m using the Sub-mm Common User 
Bolometer Array (SCUBA) on the James Clerk Maxwell Telescope (JCMT). All observations 
were performed in photometry mode. The atmospheric conditions were of exceptional quality,
with a mean $850\mu$m sky opacity of $\tau_{850}\sim0.15$. We note the possibility that sub-mm 
flux from any of the objects in our sample could in principle arise from a chance-aligned high 
redshift source, however assuming a sub-mm source density from recent blank-field sub-mm surveys 
\citep{eal,sco,fox,bor} and a conservative search radius of $2\arcsec$ (the jiggle offset on photometry mode 
observations with SCUBA) gives a probability that the flux from any one object in the sample 
comes from a chance-aligned sub-mm source of approximately 1 in 400. We therefore conclude that 
any sub-mm flux seen in our sample is most likely due to the target, rather than a background source. 
The data were reduced using the SCUBA User Reduction Facility (SURF) pipeline using standard methods.

For one host galaxy, that of SN1997ey, we obtained Hubble Space Telescope (HST) imaging from the 
HST data archive. Observations were taken using the Space Telescope Imaging Spectrograph (STIS) 
using a clear filter. The data were reduced using the IRAF reduction package {\it calstis}. The 
final image is presented in Figure 1.

\section{Results}\label{res}
The sample, their redshifts, and the sub-mm fluxes are presented in Table 1. One 
object (the host galaxy of SN 1997ey) is detected strongly at both $450\mu$m and $850\mu$m, one 
object (the host of SN2000eh) is detected at $850\mu$m only, the rest of the sample are undetected 
individually. 

As our sample is randomly selected over a narrow redshift range, we can coadd the observed sub-mm 
fluxes to measure the mean sub-mm flux  of normal galaxies at $z\sim0.5$. 
We only consider the $850\mu$m fluxes, as the $450\mu$m calibration errors are large.  The mean  observed 
$850\mu$m flux for all 17 galaxies is 1.55mJy$\pm$0.31mJy. The host galaxy of SN1997ey is however 
significantly brighter than the others, and it is probable that this source is an outlier, and that the 
flux of this galaxy will bias the mean value significantly. Assuming that the `true' source flux 
distribution is a Gaussian, then the 
distribution of fluxes for the whole sample is consistent with a gaussian at only the $\sim5\sigma$ level, 
whereas excluding the host of SN1997ey makes the distribution of fluxes consistent with a gaussian at 
$<2\sigma$, and excluding any of the remaining sources does not make the fit statistically better. We 
therefore also quote the error weighted mean flux of the 16 galaxies with the host of SN1997ey excluded,  
which is 1.01mJy$\pm0.33$mJy, and use this value in the following analysis.

\section{Discussion} \label{dis}

We first review the origin of sub-mm flux from extragalactic sources, and the relation between sub-mm 
flux and optical extinction. In many cases sub-mm emission from extragalactic sources arises in starbursts; 
compact regions ($<1$kpc across) containing $\sim10^{8}M_{\odot}$ of dust, heated by young massive 
stars forming at a rapid rate ($>50M_{\odot}$yr$^{-1}$). Since sub-mm emission measures physical dust 
volume rather than the volume of space the dust is dispersed in, the same level of sub-mm emission can 
arise from a comparable mass of dust distributed in the disk of a galaxy, heated by the general interstellar 
radiation field. This type of sub-mm emission, known as 'cirrus' emission, is seen locally, both in Galactic 
sources \citep{cas}, and in nearby galaxies \citep{bia1,dun}. Furthermore, there is evidence that 
cirrus emission may be important in blank-field sub-mm sources \citep{ef0}. As our targets contain 
a type Ia supernova, known to come from evolved stellar populations, and because none of the targets 
contain optical signatures of starbursts or AGN, we are confident that any sub-mm flux from our targets 
will be cirrus emission. 

The relation between cirrus sub-mm emission and optical extinction has been discussed by \citet{hil0}, 
\citet{cas} and \citet{bia1}. It turns out that the integrated sub-mm flux from a galaxy, with certain 
assumptions, scales linearly with $A_{V}$. We summarize these arguments here. If the dust responsible for the 
sub-mm emission is also responsible for the optical extinction, then the ratio of the extinction 
efficiency in the V band to the emission efficiency in the sub-mm at some wavelength $\lambda$ is equal 
to the ratio of the optical depths:

\begin{equation}
\frac{Q_{ext,V}}{Q_{em,\lambda}} = \frac{\tau_{V}}{\tau_{\lambda}}
\end{equation}

\noindent In the optically thin case, $\tau_{\lambda}$ can be written:

\begin{equation}
\tau_{\lambda} = \frac{I_{\lambda}}{B_{\lambda}(T)}
\end{equation}

\noindent where $I_{\lambda}$ is the sub-mm intensity and $T$ is the dust temperature. 
Solving for $A_{V}$ gives \citep{bia}:

\begin{equation}
A_{V} \propto\frac{Q_{UV}}{Q_{\lambda_{0}}}\left(\frac{\lambda}{\lambda_{0}}\right)^{\beta}\frac{I_{\lambda}}{B_{\lambda}(T)}
\end{equation}

\noindent where $\lambda$ is the observed wavelength, 
$\lambda_{0}$ is a reference wavelength, and $Q_{\lambda_{0}}$ is the extinction efficiency at the reference 
wavelength.

There are two important assumptions in this derivation, which may affect our results. Firstly, even though 
the fluxes lie within the Rayleigh-Jeans tail, the sub-mm flux still scales linearly with dust temperature, 
meaning that a mean increase in temperature could be responsible for increasing the sub-mm flux, rather than an 
increase in optical extinction. However, even assuming pure luminosity evolution between the local Universe 
and $z=0.5$, the mean dust temperature will only increase as the 1/5th power of $A_{V}$ (i.e. increasing 
with the mean surface brightness). Furthermore, the predicted rise in dust temperature between z=0 and z=0.5 
from chemical evolution models \citep{pei} is probably of the order 1K or less, insufficient to produce a 
significant rise in sub-mm flux. Secondly is the gas-to-dust ratio. When using this formula to compare optical 
extinctions between different objects, it is assumed that the gas-to-dust ratio for both objects is the same. 
When comparing local galaxies to galaxies at $z=0.5$ this criterion will not be satisfied; globally we expect 
significantly less star formation to have occurred by $z=0.5$, which may lead to a higher mean gas-to dust ratio 
than locally. Based on indirect evidence \citep{pet}, we expect this rise to be no more than $10\% - 20\%$ at 
$z=0.5$, although at higher redshifts the change may be larger \citep{pei0}. As there is no published measure 
of the gas-to dust ratio in normal galaxies at $z=0.5$, we have assumed that the change from the local mean 
value is negligible.

\subsection{Dust opacity evolution in normal galaxies} 
To establish if the mean sub-mm flux of our sample differs from the mean sub-mm flux of local normal (i.e. inactive) 
galaxies, we would like to compare to a local sub-mm survey of galaxies selected on optical flux. Currently however, 
no such survey has been published. Therefore, to estimate the expected observed mean $850\mu$m flux at $z=0.5$ 
from normal galaxies if there had been no evolution in dust opacity between the local Universe and $z=0.5$, 
we consider the local sub-millimetre survey of \citet{dun}. We have redshifted the mean $850\mu$m flux from 
this survey to z=0.5, with $k$-corrections derived using the cirrus models from \citet{ef0}. 
The resulting predicted observed $850\mu$m flux at $z=0.5$ is 0.8mJy $\pm0.1$mJy. This survey 
is however of galaxies from the IRAS Bright Galaxy Sample, which selects against galaxies with 
small dust masses, and is composed of spiral, irregular and interacting systems. The morphologies 
of the galaxies discovered by the supernova cosmology teams are however known to be $70\%$ spirals 
and irregulars, and $30\%$ ellipticals \citep{far4,sul}. As our sample is randomly selected from 
the SNeIa discovered by both the SCP and the HZT, it is reasonable to assume that the same morphological 
mix is present in our sample. In order to compare our value and the `no evolution' value, 
we must correct the `no evolution' value to reflect the morphological mix in our sample. If we assume that sub-mm 
emission from ellipticals is negligible, then this gives a predicted `no evolution' 
flux estimate at $z=0.5$ of 0.56mJy $\pm0.1$mJy. Comparing our observed mean $850\mu$m flux of 1.01mJy 
$\pm0.33$mJy (i.e. excluding the host of SN1997ey) with this value, implies a rise in sub-mm flux from 
normal galaxies at $z=0.5$ compared to locally in the range $25\% - 135\%$ ($1\sigma$ errors). 
Therefore, our results imply that sub-mm emission from normal galaxies at $z=0.5$ is comparable to or higher 
than sub-mm emission from local normal galaxies. 

We can use this mean sub-mm flux to estimate the mean value of $A_{V}$ in spiral and irregular galaxies 
at $z=0.5$. The most recent estimate of mean extinction levels in local {\it normal} galaxies (i.e including 
only an optically thin cirrus component) is that of \citet{rr2}, who derives $E(B-V)=0.1$, and therefore 
$\langle A_{V,local} \rangle = 0.31$. Hence, we obtain $\langle A_{V,z=0.5} \rangle = 0.56\pm0.17$. This 
result is in good agreement with indirect observational evidence based on the star formation history of 
the Universe \citep{lil,rr0}. It is also in agreement with model predictions for the evolution in dust 
opacity with redshift \citep{cal}, which predict that the change in mean reddening between the local 
Universe and z=0.5 will be $\Delta E(B-V)=0.05$, with a corresponding increase in extinction of $A_{V}\sim0.15$, 
and with chemical evolution models \citep{pei} which model the evolution of gas and dust mass with redshift. 

\begin{figure*}
\includegraphics[angle=0,scale=0.45]{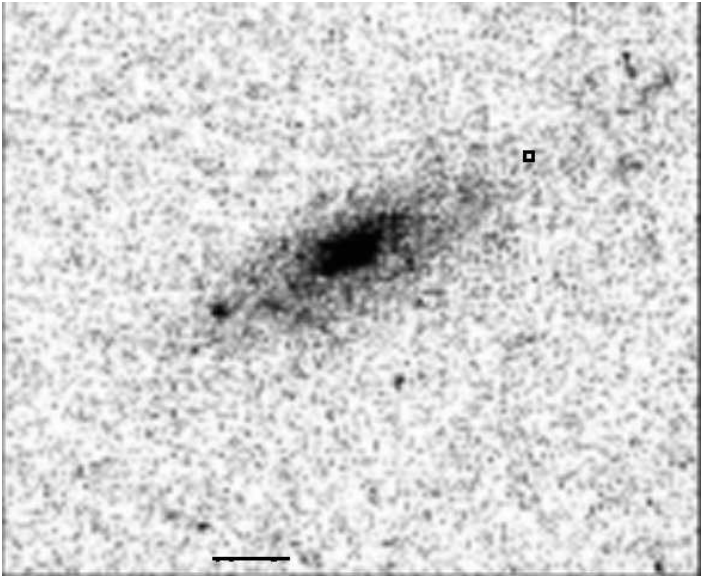}
\caption
{ 
HST STIS image of the host galaxy of SN1997ey, taken after the supernova 
had faded. The position of the supernova is marked with a square and the 
bar at the bottom denotes $1\arcsec$. \label{fig1}
}
\end{figure*}

\subsection{Dust properties in spiral galaxies at z=0.5} 
We now examine the dust properties in the host galaxy of SN1997ey, which was detected at $450\mu$m and 
$850\mu$m. One possibility is that this galaxy harbours a dust enshrouded starburst that produces the 
sub-mm flux; such a system would be unsuitable for examining the properties of dust in normal galaxies 
at $z=0.5$. This is however very unlikely, for two reasons. Firstly, as described earlier, none of the 
optical spectra of the supernova host galaxies from either the SCP or the HZT, from which this galaxy is 
selected, show signs of an AGN or starburst. Secondly, the HST STIS image for this galaxy, taken after 
the supernova had faded and presented in Figure \ref{fig1}, shows a disk galaxy with no signs of 
morphological disturbance that are seen in local \citep{sur,far1,bush} or high redshift \citep{far2} 
starburst galaxies. We therefore conclude that the observed sub-mm flux comes from the host galaxy of 
SN 1997ey, and that this galaxy is a normal disk galaxy. 

\begin{figure*}
\includegraphics[angle=90,scale=0.42]{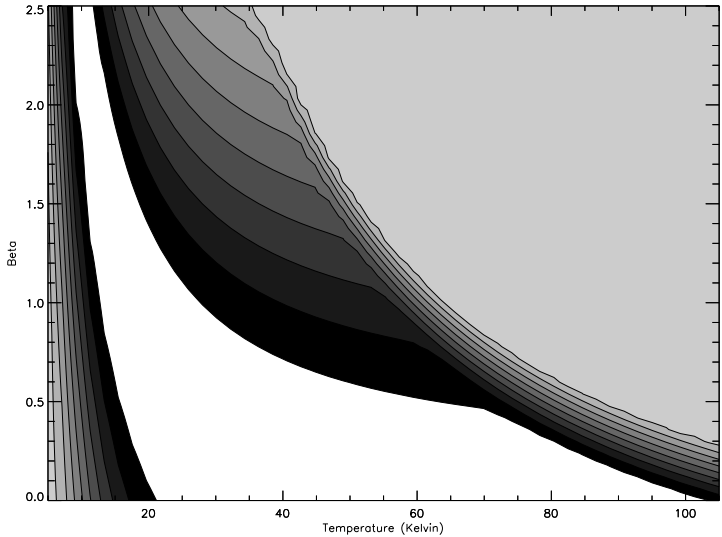}
\caption
{ 
Reduced $\chi^{2}$ distribution plot for the dust properties of the host 
galaxy of SN1997ey. The white region shows the $3\sigma$ confidence range
for dust temperature and emissivity index. Each contour then represents a 
further $3\sigma$ drop in confidence. Upper limits at $60\mu$m and $100\mu$m 
from IRAS were also used in constraining the distribution. \label{fig2}
}
\end{figure*}

As there exists sub-mm data at only two wavelengths, the properties of the dust in this galaxy are best 
estimated using a simple greybody formula. Figure \ref{fig2} shows a 
reduced $\chi^{2}$ distribution plot of $\beta$ and $T$ for the dust in the galaxy. Two particularly 
interesting cases are for $\beta\sim2$, the cirrus or `Milky-Way' type dust solution, and for $\beta\sim0$, 
the 'grey' dust solution. Whilst `grey' dust has previously been proposed to exist 
at high redshifts \citep{agui}, it is notable that, for $\beta\sim2$, the dust temperature range 
is comparable to that seen locally, and consistent with the models of \citet{pei}. We therefore find no 
observational evidence that the dust at high 
redshift differs fundamentally in nature from the dust seen in the local Universe, and conclude that 
the dust in the host galaxy of SN1997ey is normal cirrus dust.

\subsection{Normal galaxies and sub-millimeter surveys}
The detection of one object in our sample, with an $850\mu$m flux of $\sim8$mJy, highlights an alternative 
interpretation for the nature of some of the sources found in blank-field sub-mm surveys. We have observed 
17 galaxies which have $B \sim 23$, and detected one of them at an $850\mu$m flux of $\sim8$mJy.
There are approximately 3000 galaxies per sq deg at this magnitude, so our detection rate is 
$\sim200$ per sq deg. Whilst statistics based upon one object are obviously not trustworthy, it is 
interesting that this detection rate agrees surprisingly well with the detection rate reported in previous 
blank-field sub-mm surveys to comparable depths \citep{sco,fox}, who find 19 objects to 8 mJy in 260 
sq arc min, or 270 per sq deg.  

The nature of our source, and the assumed nature of blank-field survey sub-mm sources, are 
different. Blank-field survey sub-mm sources are generally thought to lie at $z > 1$, 
and to be galaxies undergoing extreme bursts of dust-shrouded star formation \citep{fox}. The 
redshift, luminosity and star formation rates are however 
in many cases estimated using only a limited set of photometric data. Our detection, which has a 
comparable flux to the sources found in these surveys, lies at $z = 0.58$ and, judging by the 
coldness of the dust and morphology of the galaxy, is forming stars at a more sedate rate. 
Whilst some sub-mm sources do now have spectroscopic redshifts that put most of them at $z\geq1$ 
\citep{lil2,cha}, it is reasonable to infer that some of the sub-mm sources in blank-field surveys are 
at lower redshifts than previously thought, and also have lower star formation rates. Whilst the 
star formation rates in these sources will be high, they still need to be carefully 
accounted for in current and future sub-mm surveys, to avoid overestimating the global history 
of star formation. 

\subsection{Implications for supernova cosmology}
There is now strong, albeit indirect, evidence from observations of the cosmic microwave background 
\citep{jaf,spe}, and of large-scale structure \citep{pea} that the total density of the Universe is 
dominated by 'dark energy'. To determine the nature of dark energy, it is necessary to track the 
evolution of dark energy with redshift directly, using a cosmological standard candle. Currently, the 
only method deployed to do this uses the luminosity distances of distant type Ia supernovae. The HZT 
and the SCP have performed surveys to find high redshift supernovae, and both groups claim that dark 
energy dominates the total density of the Universe, and that the expansion of the Universe is currently 
accelerating \citep{rie1,per1}. The HZT have further claimed \citep{rie2} to have found an earlier 
epoch of decelerating expansion, based on an apparent relative brightening of SNeIa at $z>1$  and that 
accelerating expansion commenced at $z\sim1$. This relative brightening supports a real, 'cosmological' 
influence on the supernova magnitudes, as alternative explanations (e.g. `grey' dust, progenitor evolution) 
do not readily predict this relative brightening at $z>1$. 

These results are based upon the distant supernovae being $\Delta$m = 0.40 dimmer (with a flat Universe 
prior) than if the expansion of the Universe was decelerating. There are alternative interpretations that 
can in principle account for this dimming, one of these being  host galaxy extinction. 
The issue of extinction towards distant type Ia supernovae has been examined in detail by both the SCP 
and the HZT by measuring the reddening of the supernova light curves; the SCP by comparing the rest-frame 
colours of their low and high redshift samples, and the HZT by fitting to the light-curves via the MLCS 
method. The SCP find that the reddening of both their local and distant sample, after a small number of 
reddened supernovae are removed, has a small scatter with a mean value of close to zero. 
The HZT find that only a few of their distant supernovae are significantly reddened. An analysis of the 
colour excesses derived from the supernova light curves as a function of host galaxy morphology by the 
SCP \citep{sul} also found only a modest difference in extinction between the SNe 
in ellipticals and those SNe in spirals, with $E(B-V)=-0.03$ for the SNe in the ellipticals and 
$E(B-V)=0.07$ for the SNe in spirals. Our result, of a mean value of $A_{V}$ in galaxies at $z=0.5$ that 
is $25\% - 135\%$ higher than locally, does not cast any doubt on the supernova teams results, as this 
extinction level at $z=0.5$ lies within $1.3\sigma$ of the local value, and is derived from the mean 
extinction in each galaxy rather than the line of sight extinction towards the supernovae. It does however 
highlight the need for caution in general in using supernovae as probes of the expanding Universe, as 
our derived mean extinction, $A_{V}=0.56\pm0.17$, implies a rise that is {\it at face value} comparable to the dimming 
ascribed to dark energy. Therefore, our result emphasizes the need to accurately monitor the extinction 
towards distant supernovae if they are to be used in measuring the cosmological parameters.

\section{Summary}\label{sum}

We have presented deep sub-mm observations of seventeen type Ia supernova host galaxies at $z\sim0.5$, 
from which we conclude the following:

1) The mean observed-frame sub-mm flux of the sample, excluding one bright object, is 
$S_{850}=1.01$mJy$\pm0.33$mJy. Assuming that sub-mm flux scales linearly with optical extinction, 
then this implies a rise in optical extinctions in normal, inactive galaxies compared to locally 
of $25\% - 135\%$. Scaling from the local value of $A_{V}$ given by \citet{rr2} for optically thin 
cirrus emission gives $\langle A_{V,z=0.5} \rangle = 0.56\pm0.17$. This result is in good agreement 
with both chemical evolution models \citep{pei,cal} and with other, indirect, observational 
evidence \citep{lil,rr0}. The temperature and emissivity of the dust in the brightest sub-mm 
object in our sample are comparable to the temperature and emissivity of dust in local galaxies.

2) The discovery of a moderate redshift disk galaxy with a sub-milliimetre flux comparable to the 
sources found in blank-field sub-mm surveys suggests that some of these blank-field sources 
may not be high redshift starbursts, but lower redshift dusty disks. These sources must be carefully 
accounted for in current and future sub-mm surveys, to avoid overestimating the global history of star 
formation. 

3) Our results, when combined with previous work, infer a level of extinction in galaxies at $z=0.5$ that 
could in principle produce a dimming that is comparable in size to the dimming ascribed to accelerated expansion, although 
the error on the value is large and the extinction level is the mean for the galaxy rather than the line 
of sight extinction towards the supernovae. This emphasizes the need to carefully monitor extinction levels 
towards distant supernovae if they are to be used to track the expansion rate history of the Universe.

\acknowledgments
We thank Margrethe Wold, David Branch 
and Carol Lonsdale for helpful discussion, and the referee for a very helpful report. The JCMT is 
operated by the Joint Astronomy Centre on behalf of the UK Particle Physics and Astronomy Research 
Council, The Netherlands Organization for Scientific Research and the Canadian National Research 
Council. DF was supported by NASA grant NAG 5-3370 and by the Jet Propulsion Laboratory, California 
Institute of Technology, under contract with NASA. MF and DLC were supported by PPARC. JA  
acknowledges support from the Science and Technology Foundation (FCT, Portugal) through the 
fellowship BPD-5535-2001.

\clearpage
\begin{deluxetable}{lccrrrr} 
\tablecolumns{6} 
\tablewidth{0pc} 
\tablecaption{Supernova host galaxies observed at the JCMT} 
\tablehead{ 
\colhead{Host}&\colhead{Origin$^{a}$}&\colhead{Redshift}&\colhead{$S_{450}$}&\colhead{$\sigma_{450}$}&\colhead{$S_{850}$}&\colhead{$\sigma_{850}$}}
\startdata 
SN1994al & SCP & 0.42 &   5.49 & 47.08 &  3.01 & 2.70 \\ 
SN1995aq & SCP & 0.45 &   9.52 &  9.48 & -1.33 & 1.49 \\ 
SN1995as & SCP & 0.50 &  -6.10 &  5.20 & -0.31 & 1.19 \\ 
SN1995ay & SCP & 0.48 &  -4.10 &  4.13 &  0.87 & 1.17 \\ 
SN1995az & SCP & 0.45 &   9.23 &  8.45 &  2.56 & 1.45 \\ 
SN1996cg & SCP & 0.49 &   3.46 &  3.90 & -1.11 & 1.04 \\ 
SN1997f  & SCP & 0.58 &   2.34 &  4.76 &  1.79 & 1.10 \\ 
SN1997l  & SCP & 0.55 &  -0.14 & 12.96 &  1.93 & 1.76 \\ 
SN1997em & SCP & 0.46 &   0.24 &  6.45 & -0.68 & 1.38 \\ 
SN1997ep & SCP & 0.48 &   4.12 &  4.09 &  2.01 & 1.06 \\ 
SN1997ey & SCP & 0.58 &  20.80 &  3.54 &  7.80 & 1.10 \\ 
SN1999fn & HZT & 0.47 &   1.87 &  7.04 & -0.12 & 1.28 \\ 
SN2000dz & HZT & 0.50 &  -0.51 &  8.21 &  1.87 & 1.38 \\ 
SN2000ea & HZT & 0.42 & -10.90 & 13.80 &  2.45 & 2.02 \\ 
SN2000ee & HZT & 0.47 &   1.96 &  4.63 &  1.23 & 1.06 \\ 
SN2000eg & HZT & 0.54 &   5.92 &  4.20 &  1.29 & 1.16 \\ 
SN2000eh & HZT & 0.49 &   5.18 &  6.65 &  4.61 & 1.53 \\ 
\enddata 

Fluxes are quoted in the observed frame and are given in mJy. Errors are $1\sigma$. 
$^{a}$Group responsible for discovering the supernova: `SCP' is the Supernova Cosmology Project, `HZT' 
is the High Redshift Supernova Search Team.

\end{deluxetable}

\end{document}